\documentstyle[multicol,aps,epsfig]{revtex}
\begin{document}
\newcommand{\be}{\begin{equation}}
\newcommand{\ee}{\end{equation}}
\title{Crashes, Recoveries, and `Core-shifts' in a Model of Evolving Networks}
\maketitle

\vspace{0.2truecm}
\centerline{Sanjay Jain$^{*,\dag,\ddag,\P}$ and 
Sandeep Krishna$^{*,\P}$}
\centerline{$^*${\it Centre for Theoretical Studies, 
Indian Institute of Science, Bangalore 560 012, India}} 
\centerline{$^{\dag}${\it Santa Fe Institute, 1399 Hyde Park Road, Santa Fe, NM 87501, USA}}
\centerline{$^{\ddag}${\it Jawaharlal Nehru Centre for Advanced Scientific Research,
Bangalore 560 064, India}}
\centerline{$^{\P}$ Emails: jain@cts.iisc.ernet.in, sandeep@physics.iisc.ernet.in}
\begin{abstract}
{
A model of an evolving network of interacting molecular species is shown
to exhibit repeated rounds of crashes in which several species get 
rapidly depopulated, followed by recoveries. The network inevitably
self-organizes into an autocatalytic structure, which consists of
an irreducible `core' surrounded by a parasitic `periphery'. Crashes 
typically occur when the existing autocatalytic set becomes fragile and
suffers a `core-shift', defined graph theoretically. 
The nature of the recovery after a crash, in particular the time of recovery,
depends upon the organizational structure that survives the crash.
The largest eigenvalue of the adjacency matrix of the graph is an
important signal of network fragility or robustness. 
}
\end{abstract}

\begin{multicols}{2}

The dynamics of crashes and recoveries
has been the subject of several empirical and modeling studies in 
macroevolution 
(for reviews, see \cite{NP,Drossel}) and finance \cite{MS,Bouchaud,JS}.
The main attempt of most models has been to reproduce quantitatively
the observed statistics of event sizes. However, it is also worthwhile
to ask whether large events share other common features, or signatures,
that precede the event or characterize the kind of systemic transformation
caused by them. Here we present a structural analysis, based on 
network properties, of events that occur in a model \cite{JK1}\
in which populations of molecular
species co-evolve with their network of catalytic interactions
\cite{BFF,Kauffman1,FB}. We find large crashes to be associated 
with a particular kind of structural change in the network, which we
call a `core-shift', and identify network characteristics that
signal the system's susceptibility to crashes. This kind of
analysis might be useful for other models of biological and social
evolution.  

The system is a directed graph with $s$ nodes labeled 
$i\in S\equiv\{1, 2, \ldots, s\}$, represented by its adjacency matrix
$C\equiv(c_{ij})$. If there exists a directed link from node $j$ to node $i$ in the graph
then $c_{ij}=1$, else $c_{ij}=0$. Each node represents a molecular
species in a prebiotic pond and $c_{ij}=1$ means that $j$ is a catalyst
for the production of $i$.
The dynamical variables are the `relative population vector' of 
the species ${\bf x} \equiv \{ (x_1,\ldots,x_s)|0 \le x_i \le 1,
\sum_{i=1}^{s}x_i=1 \}$, which is a fast variable, and the graph itself
(or $C$), which is a slow variable.
Initially, each $c_{ij}$ for $i\ne j$ is independently
chosen to be unity with a probability $p$ and
zero with a probability $1-p$. To exclude self-replicating species, 
$c_{ii} \equiv 0$ for all $i$. 
Each $x_i$ is chosen randomly in $[0,1]$ and all $x_i$ are rescaled
so that $\sum_{i=1}^{s}x_i=1$.
With $C$ fixed, ${\bf x}$ is evolved according to 
\be
\dot{x}_i = \sum_{j=1}^s c_{ij} x_j -  x_i \sum_{k,j=1}^s c_{kj} x_j
\label{xdot}
\ee
until it reaches its attractor (always a fixed point \cite{JK1,JK2}), denoted ${\bf X}$. 
(\ref{xdot}) is an idealization of rate equations for catalyzed
reactions in a well stirred chemical reactor.
The set $\cal{L}$ of nodes with the least
$X_i$ is determined, 
i.e, ${\cal{L}}=\{i \in S|X_i = {min}_{j \in S} X_j \}$.
A node, denoted $k$, is picked randomly from $\cal{L}$ and for every $i\ne k$
$c_{ik}$  and $c_{ki}$ are independently
reassigned to unity with probability $p$ and zero with probability $1-p$, irrespective
of their earlier values. This corresponds to removing the node $k$ and all its links
from the graph and replacing it by a new node $k$ with random links to and from the other nodes. $c_{kk}$ is
set to zero, $x_k$ is set to a small constant $x_0$, all other $x_i$ are perturbed by a small amount 
from their existing value $X_i$, and all $x_i$ are rescaled so that $\sum_{i=1}^s x_i=1$.
This captures, in an idealized way, the impact
of a periodic fluctuation like a tide or flood, which can wash out one of
the least populated species in the pond (extremal selection \cite{BS}), and
bring in a new molecular species whose catalytic links with those in the
pond are random (introduction of novelty).
Then ${\bf x}$ is again evolved to its new attractor, another graph update is performed, 
and so on.

Fig. 1 shows the number of populated species in the attractor (i.e., species with $X_i>0$), $s_1$, as a function of time for
three runs with different $p$ values. 
Time is represented by $n$, the number of graph updates. 
Three regimes or phases of behaviour can be observed. First,
the `random phase'
in which $s_1$ fluctuates about a low value. 
Second, the `growth
phase', when $s_1$ shows a clear rising tendency (occasionally
punctuated by drops). Third, the `organized phase' where
$s_1$ stays close to its maximum value, $s$. The average time spent in each
phase depends upon $p$ and $s$.
In this letter we investigate the 
large and sudden drops in $s_1$, visible in Fig. 1 (mentioned briefly in
\cite{JK3}).
These `crashes' in the organized and growth phases
are followed by `recoveries', in which $s_1$ rises on a certain timescale. 
Fig. 2 shows the probability distribution $P(\Delta s_1)$ of
changes in the number of populated species, $\Delta s_1(n) \equiv 
s_1(n) - s_1(n-1)$. The asymmetry between rises and drops as well
as fat tails in the distribution of fluctuations are evident. 
For low $p$ the probability of large drops is an order
of magnitude greater than intermediate size drops (also see Fig. 1) 
\cite{earthquake}. 

An autocatalytic set (ACS) is said to be a set of species 
which contains a catalyst for each of its 
members 
\cite{Eigen,Kauffman2,Rossler}. 
Here it is
a subgraph each of whose nodes has at least one incoming link
from a node of the same subgraph. (A subgraph is a subset of nodes
together with all their mutual links.) 
For example in Fig. 3, the subgraph formed by nodes 40, 93, 36, 51 is not an ACS, but
that formed by 40, 93, 36, 51, 63 is. The subgraph of all
black nodes is also an ACS.
Let $\lambda_1(C)\equiv\lambda_1$ be the largest eigenvalue of $C$. It can
be shown 
\cite{JK2}
that ({\it i}) if the graph does not have an ACS then $\lambda_1=0$, and if it does 
then $\lambda_1 \geq 1$.
({\it ii}) ${\bf X}$ is an eigenvector of $C$ with
eigenvalue $\lambda_1$. ({\it iii}) The set of
nodes for which $X_i > 0$ is uniquely determined by $C$, independent of 
(generic) initial condition on ${\bf x}$. ({\it iv}) 
If $\lambda_1\ge 1$ the subgraph formed by the set in ({\it iii}) 
constitutes an ACS,
which will be referred to as the `dominant ACS'.
(We find {\bf X} from these algebraic properties, rather than 
numerically integrating (\ref{xdot}).)

In the random phase the graph has no ACS (in Fig. 1 this phase coincides with
$\lambda_1=0$).
The graph remains random 
because non ACS structures are not robust \cite{JK3}.
This phase continues on average for a time $\tau_a=1/p^2s$ until at some graph update 
a small ACS appears by chance and the growth phase begins (in Fig. 1, $\lambda_1$
jumps from zero to one at that very time step, and in general, at the 
beginning of every growth phase). 
A small ACS is robust. ({\it iv}) implies that
members of the ACS do well populationally
compared to species outside it, 
hence the latter are replaced in subsequent graph updates. When the new
species receives a link from the existing dominant ACS, the latter typically
expands and $s_1$ increases. This growth and self-organization 
continues over a timescale $\tau_g \ln s$ where $\tau_g=1/p$ until the
dominant ACS spans the entire graph and $s_1$ becomes equal
to $s$ (see Fig. 3) \cite{JK1,JK2,JK3}. That marks the beginning of 
the organized phase. Note that the entire graph in Fig. 3 is an ACS.
In a fully spanned ACS the least populated species must be a member of
the ACS. Now competition between members of the dominant ACS becomes 
important and can lead to fragility and rupture of the organization. 

Let us define a {\it crash} as a graph update event $n$ for which
$\Delta s_1(n) < -s/2$, i.e., an event in which a significant number
(arbitrarily chosen as $s/2$) of the species go extinct.
In runs with $s=100, p=0.0025$ totaling 1.55 million iterations we observed
701 crashes. 
It is evident from Fig. 1 that crashes typically take place at or near $\lambda_1=1$.
This can be understood by taking a closer look at the structure of the dominant ACS.

The dominant ACS consists of a `core' and a `periphery'.
The {\it core} of a dominant ACS is the maximal subgraph, $Q$, 
from each of whose nodes all
nodes of the dominant ACS can be reached along some directed
path. The rest of the dominant ACS is its {\it periphery}. For an example
see Fig. 3. 
When the dominant ACS consists of two
or more disjoint subgraphs the above definition
applies to each component separately \cite{definition}. 
This distinction between 
core and periphery is useful in the context of the above dynamics. 
For example, the ratios of $X_i$ values of the
core nodes are unchanged if any periphery node or link is removed from the
dominant ACS, but removing or adding
any node or link to the core in general changes all $X_i$ ratios.
For any subgraph $A$ define $\lambda_1(A)$ to be the largest eigenvalue of the
submatrix of $C$ corresponding to $A$. Then, 
it can be shown that $\lambda_1(Q)$ is the same as 
the largest eigenvalue of the whole graph, $\lambda_1$.
The core (of each component) is an irreducible subgraph (i.e., one
which contains at least two nodes and 
a directed path from each of its nodes to each of its other nodes).
It follows from the Perron-Frobenius theorem that if some links are added
to the core (with possibly additional nodes) $\lambda_1$ increases,
and if removed from the core, $\lambda_1$ decreases. 
Thus $\lambda_1$ is a
measure of the core size and multiplicity of pathways or `redundancy'
within it. 
$\lambda_1=1$ corresponds to the case where the core
(of every disjoint component of the dominant ACS)
has exactly one cycle. Such a core has no internal
redundancy; the removal of any link from it will cause the ACS
property (of that component) to disappear.

This is one, purely graph theoretical, reason why the organization is 
fragile in the vicinity of $\lambda_1=1$. Another reason is dynamical: 
when $\lambda_1>1$ the core nodes are better protected against selection
by virtue of their larger populations, whereas at $\lambda_1=1$ they are 
more vulnerable.
The reason is as follows:  Since ${\bf X}$ is an eigenvector of $C$
with eigenvalue $\lambda_1$, when $\lambda_1 \neq 0$
it follows that for nodes that belong to the dominant ACS,
$X_i=(1/{\lambda_1})\sum_j c_{ij}X_j$.
In particular, if a node $i$ of the dominant ACS has only one incoming link
(from the node $j$, say)
then $X_i=X_j/\lambda_1$, i.e., $X_i$ is `attenuated' with respect to
$X_j$ by a factor $\lambda_1$.
The periphery of an ACS is a tree like structure emanating
from the core, with most nodes having a single incoming link, for small $p$.
Consider for example Fig. 3, in which the entire graph is an ACS with $\lambda_1=1.31$, 
and focus in particular on the chain
of nodes $44\rightarrow 45\rightarrow 24\rightarrow 29\rightarrow 52\rightarrow 89\rightarrow 86\rightarrow 54\rightarrow 78$.
The farther down such a chain a periphery node is, the lower is its
$X_i$ because of the cumulative attenuation. For such an ACS with $\lambda_1>1$
the `leaves' of the periphery tree will typically be the species with
least $X_i$ (and node 78 in Fig. 3 is one such). 
However, when $\lambda_1=1$
there is no attenuation. Periphery nodes will not have
lower $X_i$ than core nodes and some may have higher if they have more than
one incoming link. Thus at $\lambda_1=1$ the core is not protected and in fact
will always belong to $\cal{L}$
if the ACS spans the graph. $\lambda_1$ is known to be of 
significance in other complex systems as well \cite{LCBM,PGRAS,FDBV,GKK}. 

We now present evidence that crashes are indeed due to changes in the 
structure of the core.
Define the {\it core overlap}, denoted $Ov(C,C')$, between two graphs $C$ and $C'$
(whose nodes are labeled) as the number of common links in the cores $Q$ and 
$Q'$ of their dominant ACSs (i.e., 
the number of ordered pairs of nodes $(i,j)$ such that
$Q_{ij}$ and $Q'_{ij}$ are both non-zero.) If either $C$ or $C'$ does not have an ACS,
$Ov(C,C')$ is by definition zero. A graph update event at time $n$ will be called a
{\it core-shift} if $Ov(C_{n-1},C_n)=0$ ($C_n$ is the graph at time $n$).
Fig. 4 shows that most (612) of the 701 crashes
were core-shifts.
(If a crash is defined as an event in which more than $90\%$ of 
the species become extinct, then there are 235 crashes in these runs of which
226 are core-shifts.) Of the remaining 89 crashes 79 were `partial
core-shifts' and 10 were events in which the core remained unchanged.

In the 612 core-shifts, the average number of incoming plus outgoing links 
is 2.27 for all nodes in the graph,
2.25 for the node that is hit and 1.25 for the new node.
Thus the nodes whose exchange causes the
crash are not excessively rich in links (and the hit node 
is always the least populated). `Nondescript' nodes
such as these cause system wide crashes because of their
critical location in a
small core (the average core size at the 612 core-shifts is 6.3
nodes) that is responsible for the coherence and sustenance of the whole
network \cite{JK5}. 
Core-shifts in which the ACS
is completely destroyed typically cause the
largest damage (of 612 core-shifts these are 136 in number, with 
$|\Delta s_1| = 98.2 \pm 1.2$). The remaining 476 in which an ACS
exists after the core-shift have $|\Delta s_1| = 75.0 \pm 14.2$. 
The former constitute an increasing fraction of the crashes
at smaller $p$ values, causing the upturn in $P(\Delta s_1)$
at large negative $\Delta s_1$ for small $p$ (Fig. 2).

Let $\tau_s$ denote the time for which the system stays in the organized phase 
until a core-shift occurs.  $\tau_s$ increases with $p$ but 
its quantitative dependence on $p$ and $s$ remains an open question.
After a crash if there is no ACS the graph usually becomes a random
graph in order $s$ time steps. It takes on average $\tau_a=1/p^2s$
time steps before a new ACS forms \cite{JK1}. Once an ACS appears
it grows exponentially across the graph on a time scale $\tau_g
= 1/p$.
After crashes in which an ACS survives 
the recovery time scale is just $1/p$.
The asymmetry between positive and negative changes in $s_1$ is a 
natural consequence of different processes being involved in the two cases.

It is characteristic of natural evolution that
as different structures arise in the system the nature of the selective pressure
on existing structures, and hence their effective dynamics, changes. 
In the present system we likewise see an effectively
random graph evolution when there is no ACS, a self-organizing
growth phase when an ACS, a small cooperative and hence robust
structure, arises, and competition within the ACS resulting in its
eventual fragility
when a fully autocatalytic
graph is formed. 
A different system in which the selective pressures and dynamics change as new
structures arise is discussed in
\cite{CRA}.
Robust yet fragile structures also arise in
highly designed systems
\cite{CD}. In the present model
the appearance of different structures dynamically generates different time 
scales:
$\tau_a$ in the random phase, $\tau_g$ in the growth phase, and
the survival time of the core, $\tau_s$, in the organized phase.
These multiple structures and timescales arise endogenously, i.e., 
they are all consequences of the same underlying
dynamical rules.

We thank O. Narayan for a discussion and drawing our attention to 
\cite{earthquake}.
S.J. acknowledges the Associateship of Abdus Salam International
Centre for Theoretical Physics, Trieste, and hospitality of the Max Planck Institute
for Mathematics in the Sciences, Leipzig.
This work was supported in part by a grant from
the Department of Science and Technology, Government of India.

\noindent {\bf Figure legends\\} 
\noindent
{\bf Figure 1.} The number of populated species, $s_1$ (continuous line), and
the largest eigenvalue of $C$ (whose significance is discussed later in the text),
$\lambda_1$ (dotted line), versus time, 
$n$. The $\lambda_1$ values shown are $100$ times the actual $\lambda_1$ value. Runs shown have $s=100$,
and {\bf (a)} $p=0.001$, {\bf (b)} $p=0.0025$ and {\bf (c)} $p=0.005$. 

\noindent
{\bf Figure 2.} Probability distribution of changes in the number of populated
species. $P(\Delta s_1)$ is the fraction of time steps in which $s_1$
changes by an amount $\Delta s_1$ in one time step in an ensemble of runs with $s=100$ and $p=0.001,0.0025,0.005$.
Only time steps where an autocatalytic set initially exists are counted.

\noindent
{\bf Figure 3.} The structure of the graph at $n=2885$ 
for the run in Fig. 1b, when the 
dominant ACS spanned the entire graph for the first time.
Node numbers $i$ from $1$ to $100$
are shown in the circles representing the nodes. 
Black circles correspond to nodes in the `core' of the ACS,
and grey to the `periphery', defined in the text.

\noindent
{\bf Figure 4.} 
Frequency, $f$, of core overlaps in crashes for runs with
$s=100, p=0.0025$. The $x$-axis displays the value of
$Ov(C_{n-1},C_n)$ in crashes (i.e., in the 701 events with $\Delta s_1(n) 
<-50$).\\

\begin{figure}
\epsfysize=5cm
\epsfxsize=8cm
\noindent
\epsfbox{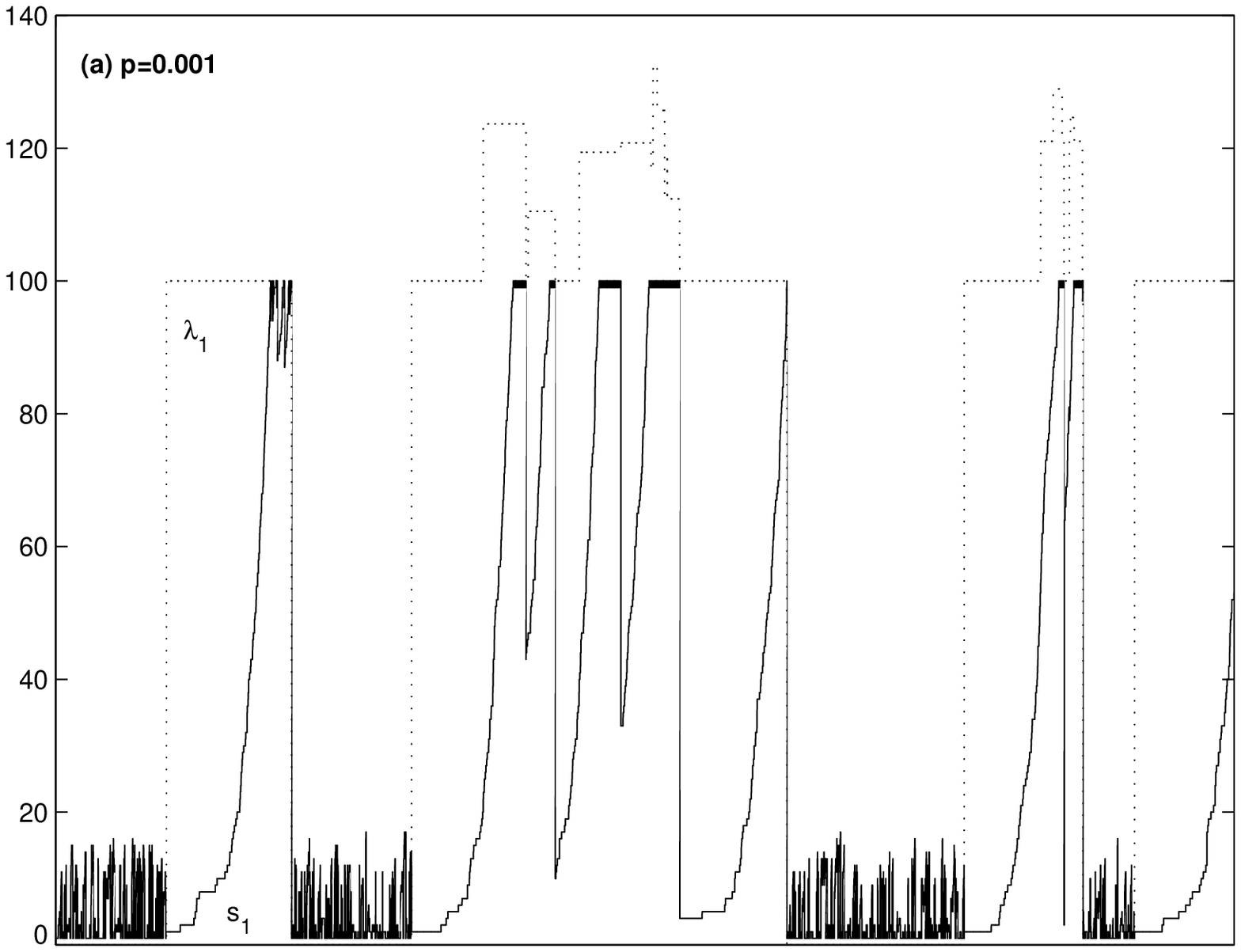}
\end{figure}
\begin{figure}
\vspace{-1cm}
\epsfysize=5cm
\epsfxsize=8cm
\noindent
\epsfbox{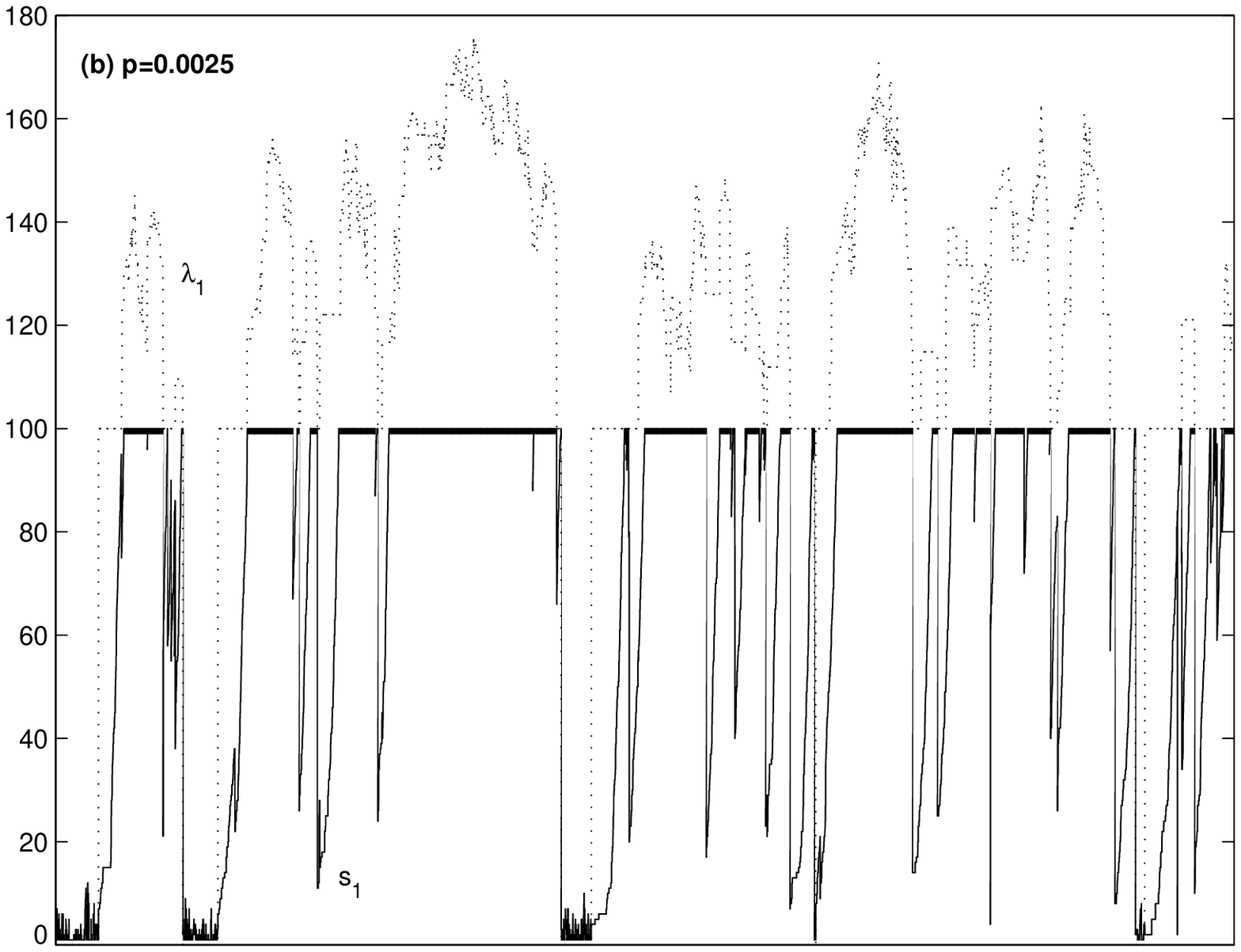}
\end{figure}
\begin{figure}
\vspace{-1cm}
\epsfysize=5cm
\epsfxsize=8cm
\noindent
\epsfbox{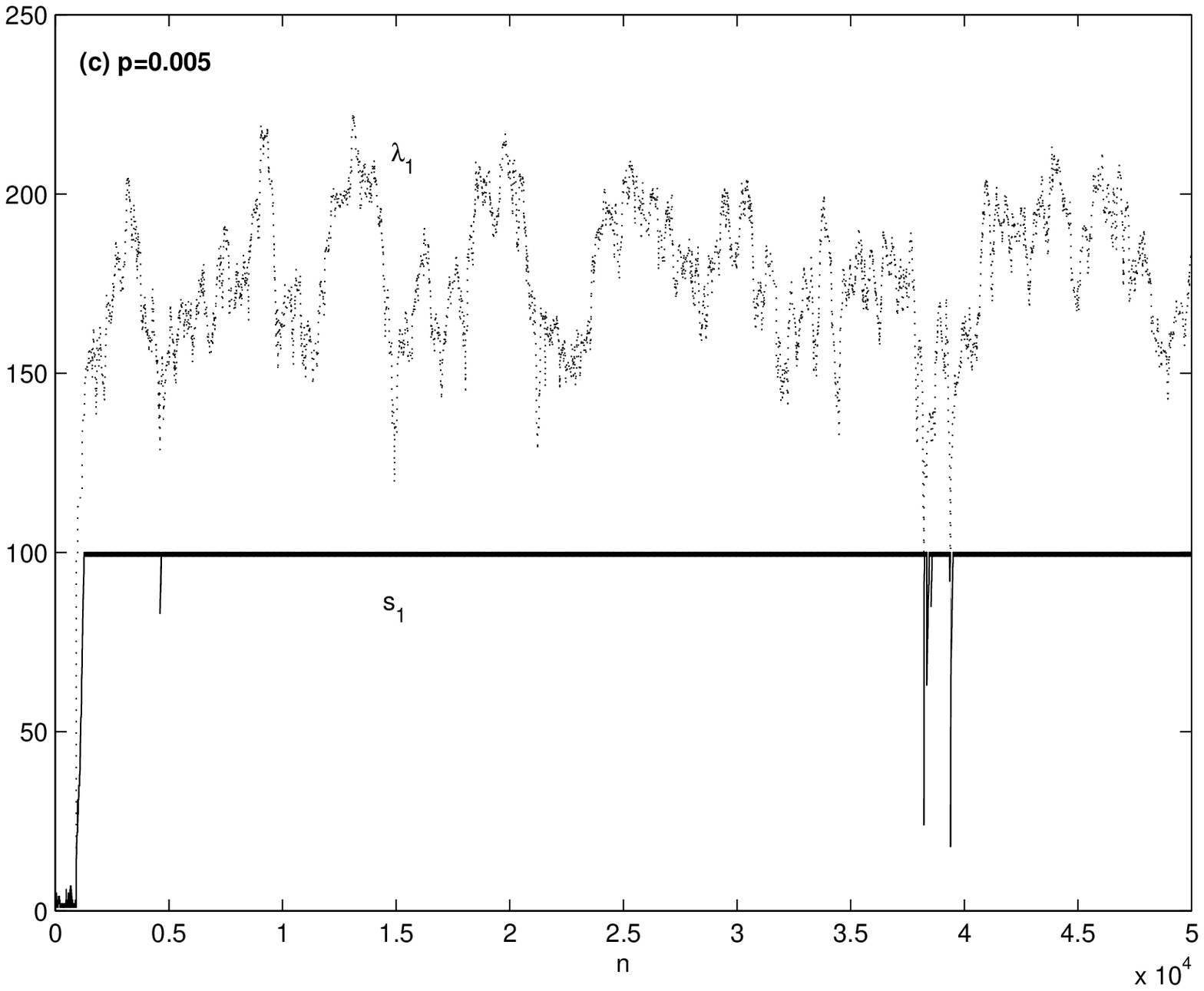}
{\bf Figure 1.} 
\end{figure}

\noindent
\begin{figure}
\epsfxsize=8cm
\noindent
\epsfbox{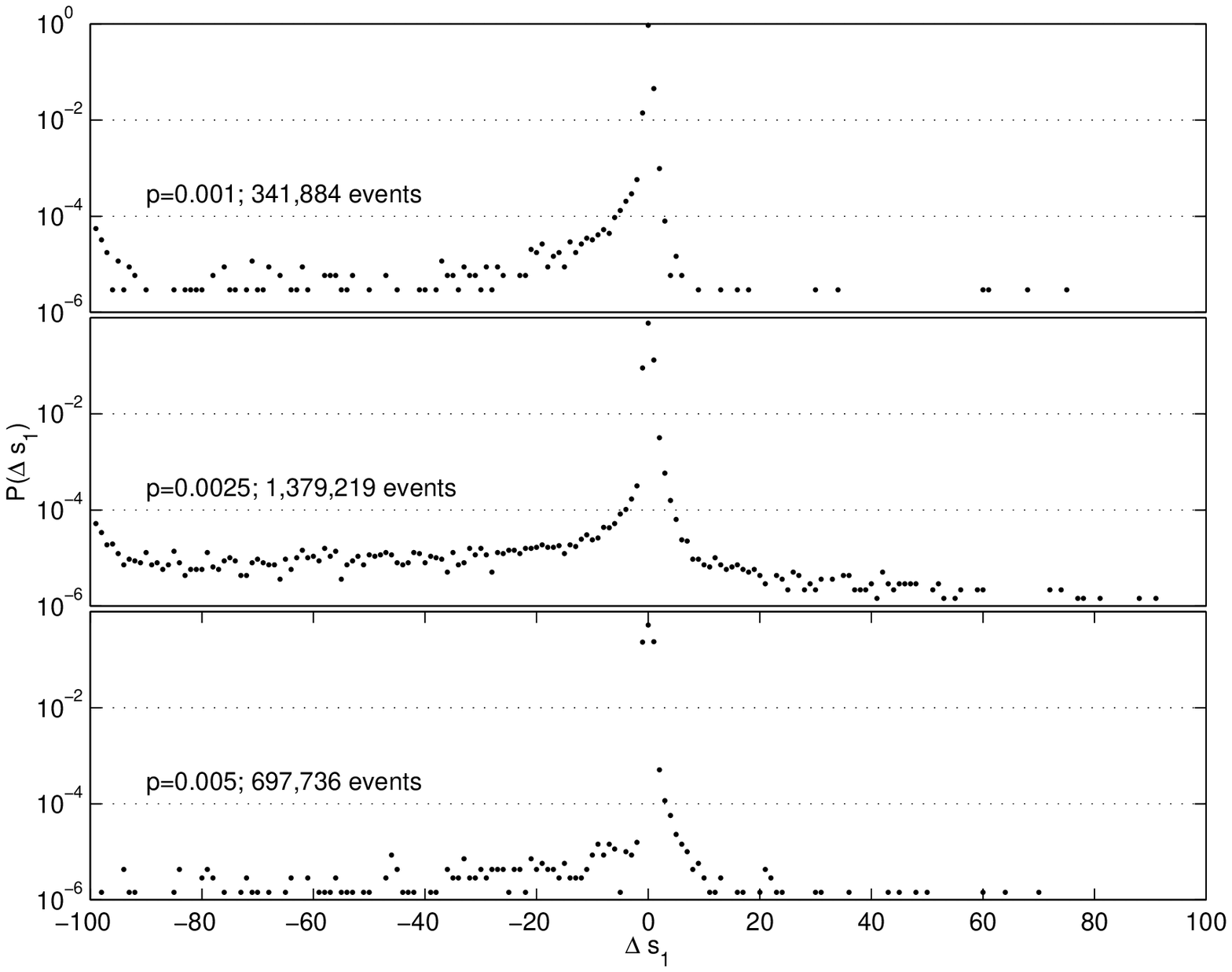}
\end{figure}

\noindent
\vspace{-1cm}
{\bf Figure 2.} 

\begin{figure}
\epsfysize=8cm
\noindent
\epsfbox{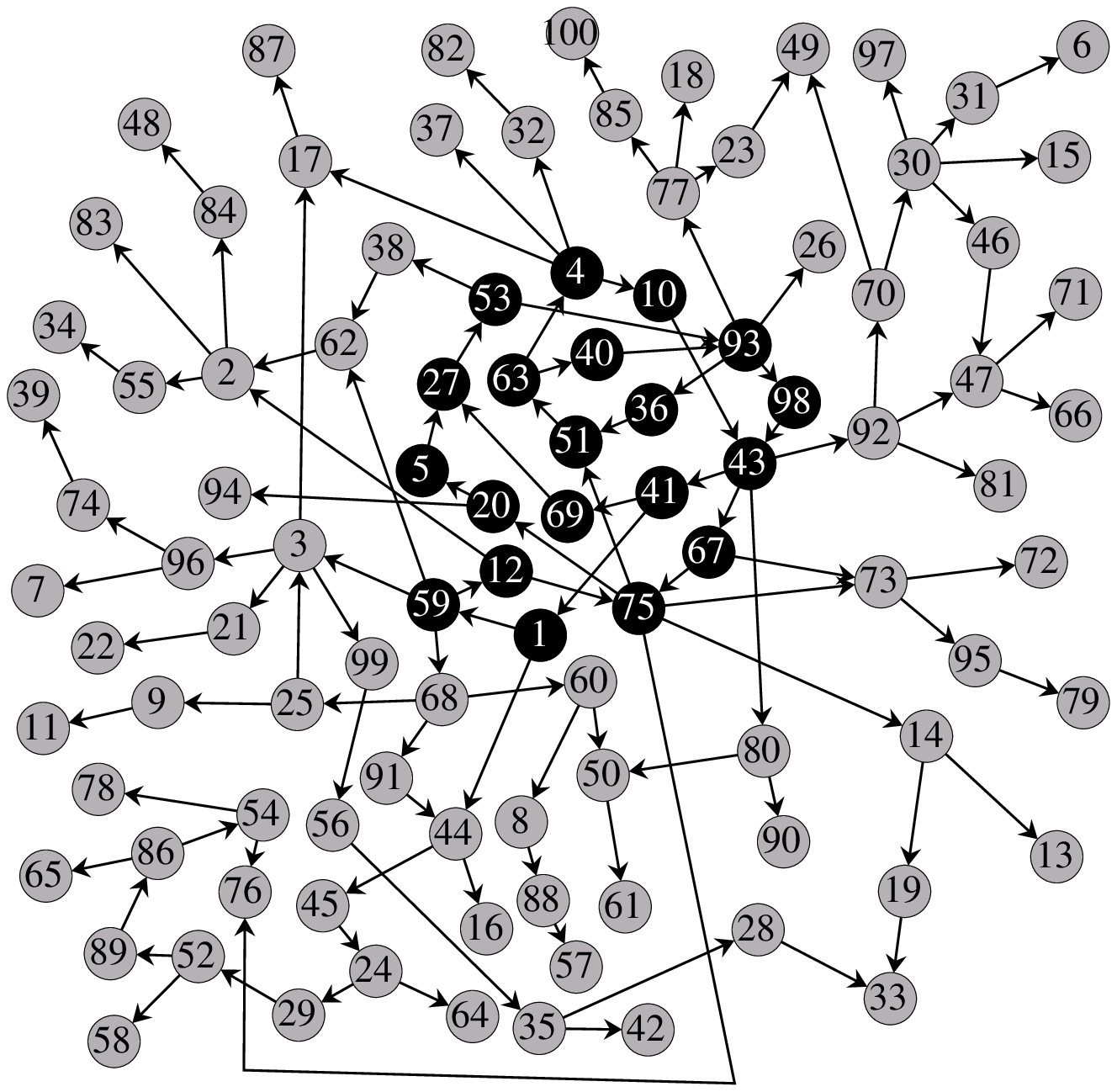}
\end{figure}

\vspace{-1cm}
{\bf Figure 3.} 
\noindent
\begin{figure}
\epsfxsize=8cm
\noindent
\epsfbox{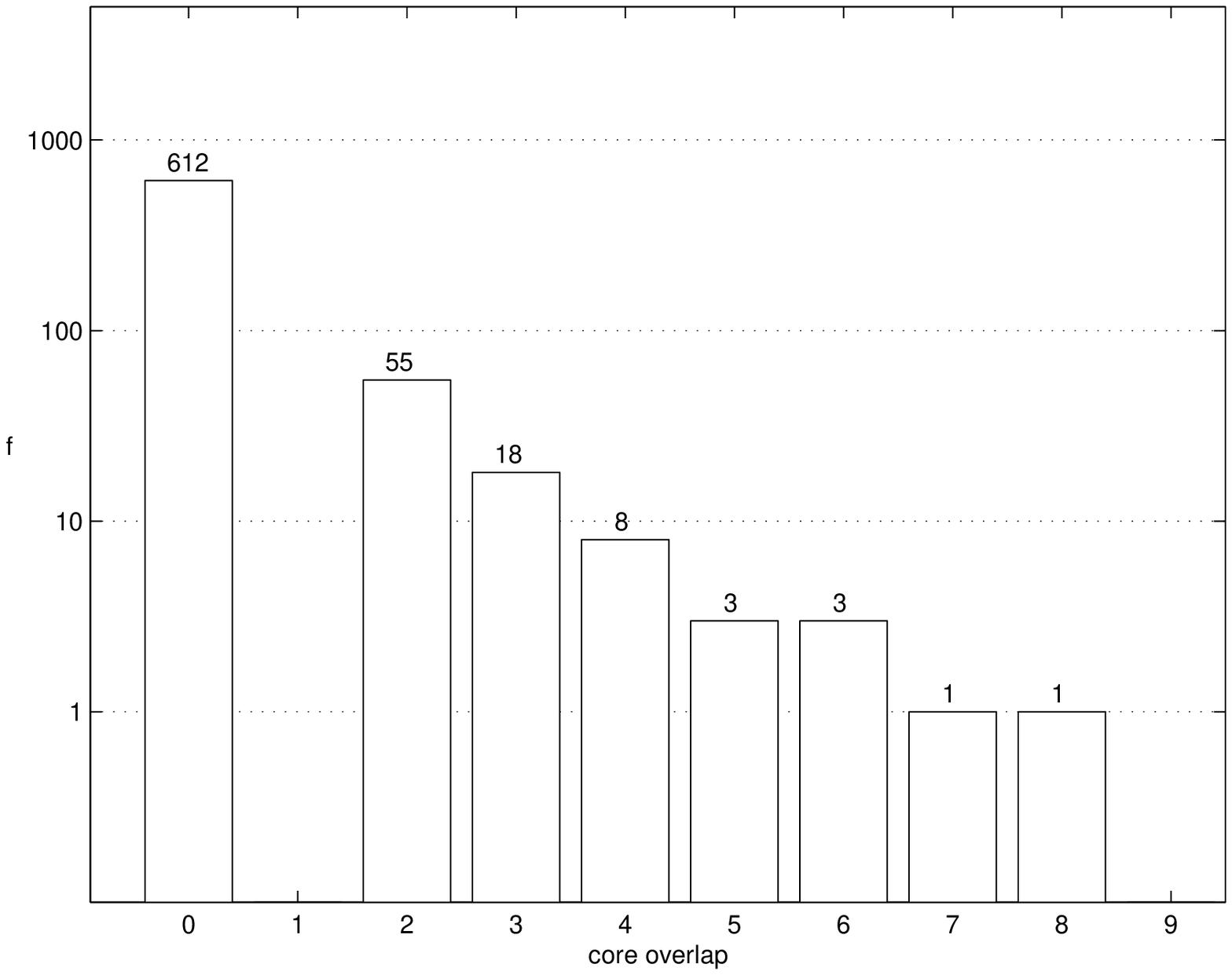}
\end{figure}

\noindent
\vspace{-1cm}
{\bf Figure 4.} 

\end{multicols}

\begin{thebibliography}{99}
\bibitem{definition}{Some rarely occurring ACS structures which do not
alter our main conclusions need a more general definition of core and
periphery (to be discussed elsewhere).}
\end{thebibliography}
\end{document}